# *Hybrid Nanophotonics*

Sergey Lepeshov[1], Alexander Krasnok[1,2*], Pavel Belov[1], Andrey Miroshnichenko[3]

[1]*Laboratory of Nanophotonics and Metamaterials, ITMO University, St. Petersburg, Russia*

[2]*Department of Electrical and Computer Engineering, The University of Texas at Austin, Austin, TX, 78712, USA*

[3]*Nonlinear Physics Centre, Australian National University, Canberra ACT 2601, Australia*

E-mail: akrasnok@utexas.edu

## Introduction

*Advances in the field of plasmonics, that is, nanophotonics based on optical properties of metal nanostructures, paved the way for the development of ultrasensitive biological sensors and other devices whose operating principles are based on localization of an electromagnetic field at the nanometer scale. However, high dissipative losses of metal nanostructures limit their performance in many modern areas, including metasurfaces, metamaterials, and optical interconnections, which required the development of new devices that combine them with high refractive index dielectric nanoparticles. Resulting metal-dielectric (hybrid) nanostructures demonstrated many superior properties from the point of view of practical application, including moderate dissipative losses, resonant optical magnetic response, strong nonlinear optical properties, which made the development in this field the vanguard of the modern light science. This review is devoted to the current state of theoretical and experimental studies of hybrid metal-dielectric nanoantennas and nanostructures based on them, capable of selective scattering light waves, amplifying and transmitting optical signals in the desired direction, controlling the propagation of such signals, and generating optical harmonics.*

# *Гибридная Нанофотоника*

Сергей Лепешов[1], Александр Краснок[1,2*], Павел Белов[1], Андрей Мирошниченко[3]

[1]*Laboratory of Nanophotonics and Metamaterials, ITMO University, St. Petersburg, Russia*

[2]*Department of Electrical and Computer Engineering, The University of Texas at Austin, Austin, TX, 78712, USA*

[3]*Nonlinear Physics Centre, Australian National University, Canberra ACT 2601, Australia*

E-mail: akrasnok@utexas.edu

## Абстракт

*Успехи в области плазмоники, то есть нанофотоники основанной на оптических свойствах металлических наноструктур, проложили путь к разработке сверхчувствительных датчиков, биологических сенсоров и других устройств, принцип работы которых основан на локализации электромагнитного поля на нанометровых масштабах. Однако высокие тепловые потери металлических наноструктур ограничивают их использование во многих современных областях, включая метаповерхности, метаматериалы, и нановолноводы, что потребовало разработки новых устройств, сочетающих их с диэлектрическими наночастицами с высоким показателем преломления. Такие металло-диэлектрические (гибридные) наноструктуры продемонстрировали много интересных с точки зрения практического применения свойств, включая низкие тепловые потери, оптический магнитный резонансный отклик, сильные нелинейно-оптические свойства, что сделало разработки в данной области авангардом современной науки о свете. Данный обзор посвящен современному состоянию теоретических и экспериментальных исследований гибридных металлодиэлектрических наноантенн и наноструктур на их основе, обладающих способностью избирательно рассеивать световые волны, усиливать и передавать в*

*заданном направлении оптические сигналы, управлять распространением таких сигналов и генерировать оптические гармоники.*

**Ключевые слова**: Нанофотоника, плазмоника, высокоиндексные диэлектрические наночастицы, оптический магнитный отклик, гибридные наноструктуры, наноантенны

## 1. Введение

Управление светом на нанометровых масштабах является приоритетным направлением исследований как в фотонной и информационной индустриях, так и в науках о жизни. Многочисленными исследователями по всему миру было обнаружено, что управлять электрической и магнитной компонентами оптической волны на нанометровом масштабе можно, используя резонансные наноструктуры. Такие наноструктуры, являющиеся основой современной нанофотоники, получили название *оптические наноантенны* [1–12]. Они обладают огромным прикладным потенциалом в различных областях: от фотовольтаики [13–17] и оптической обработки информации [2,5,18–20] до микроскопии со сверхразрешением в биологии [21–25] и терапии онкологических заболеваний в медицине [26–30]. Недавние исследования показали, что наноантенны позволяют создавать источники одиночных фотонов [31–35], нанолазеры [36–40], а также эффективные преобразователи видимого и ближнего инфракрасного излучения в среднее и дальнее инфракрасное [41,42]. Такой широкий спектр практических применений наноантенн и наноструктур, состоящих из них, обеспечивается за счет их способности к эффективному преобразованию свободно распространяющихся волн в ближнее электромагнитное поле и многократному увеличению интенсивности поля в субволновых областях [43–45]. Кроме того, было показано, что оптические наноантенны могут увеличить скорость спонтанной эмиссии оптического излучения квантового источника (эффект Перселла) [46–53], а также сконцентрировать и перенаправить это излучение в заданную область пространства [54,55].

Выделяют несколько видов наноантенн различающихся типом материалов, использующихся для их изготовления, и как следствие, физическими принципами работы: металлические, диэлектрические и металло-диэлектрические (гибридные). Металлические наноантенны, зачастую называемые плазмонными, характеризуются сильной локализацией возбуждающего электромагнитного поля в масштабах порядка 100 нм и демонстрируют высокие значения фактора Перселла, благодаря возбуждению в них локализованного плазмонного резонанса [56]. Традиционно для их производства применяются химически стойкие благородные металлы - золото или серебро, имеющие хотя и наименьшие среди металлов, но все же высокие диссипативные потери в видимом диапазоне [57]. Это ограничивает использование таких наноструктур в системах, где требуется эффективная передача оптической энергии, однако, открывает новые горизонты для приложений в термоплазмонике и медицине [58].

Преодолеть ограничения плазмонных наноантенн призваны диэлектрические наноантенны [59]. Диэлектрические наноантенны изготавливаются из материалов с высоким показателем преломления и как можно меньшим коэффициентом поглощения в видимом диапазоне [60]. Поскольку они сделаны из диэлектриков допускающих свободное проникновение электромагнитного поля внутрь наноструктур, такие наноантенны обладают как электрическим, так и магнитным резонансами Ми [8,52,61–77]. Эти резонансы позволяют наноантеннам взаимодействовать с электрической и магнитной компонентами волны в равной степени, что делает диэлектрические наноантенны достойной альтернативой плазмонным, оперирующими лишь электрической составляющей света [78]. Тем не менее, диэлектрические наноантенны имеют недостаточно высокую степень локализации поля и, как следствие, меньший фактор Перселла, чем плазмонные.

В последнее время ведутся разработки гибридных гибридных наноструктур, объединяющих в себе преимущества металлических и диэлектрических наноантенн, способных выполнять эффективное преобразование падающего поля в ближнее и обратно, и управлять магнитной компонентой. Было показано, что такие гибридные наноструктуры обладают многими, интересными с точки зрения практического применения свойствами, включая низкие тепловые потери, оптический магнитный резонансный отклик, сильные нелинейно-оптические свойства, что сделало разработки в данной области авангардом современной науки о свете. Данный обзор посвящен современному состоянию теоретических и экспериментальных исследований гибридных металлодиэлектрических наноантенн и наноструктур на их основе, обладающих способностью избирательно рассеивать световые волны, усиливать и передавать в заданном направлении оптические сигналы, управлять распространением таких сигналов и генерировать высшие гармоники. Отметим, что в настоящее время не существует обзорных работ в области гибридной нанофотоники не только в отечественной, но и в мировой литературе, что делает работу важной для данной области исследования. Работа построена следующим образом. В Главе 2 мы обсуждаем оптические свойства плазмонных и диэлектрических наночастиц. В Главе 3 обсуждаются работы в области гибридных наноструктур для управления оптическим излучением квантовых источников. В Главе 4 приводятся общие сведения из теории эффекта Перселла и приводятся результаты исследования влияния гибридных наноструктур на скорость спонтанного излучения квантовых источников. Наконец, в Главе 5 обсуждаются нелинейные и перестраиваемые гибридные наноструктуры.

## 2. Оптические свойства плазмонных и диэлектрических наночастиц

Оптический отклик плазмонных (металлических) и диэлектрических наночастиц сферической формы может быть рассмотрен с единой точки зрения основанной на точном решении соответствующей задачи дифракции, впервые полученном Густавом Ми в 1908 году [61]. Мы приводим здесь некоторые общие выводы, следующие из этой теоретической модели. Поле плоской волны рассеянной на сферической частице может быть записано в виде суммы парциальных сферических волн с коэффициентами $a_l$ и $b_l$. Таким образом, могут быть записаны выражения для нормированного полного сечения рассеяния ($Q_{scs}$), а так же сечения поглощения ($Q_{abs}$) и сечения экстинкции ($Q_{ext}$) для сферической частицы любого радиуса $R$ расположенной в пространстве, заполненном немагнитным диэлектриком с диэлектрической проницаемостью $\varepsilon_h = n_h^2$ в следующем виде [79,80]:

$$Q_{scs} = \frac{2}{(kR)^2} \sum_{l=1}^{\infty} (2l+1)(|a_l|^2 + |b_l|^2),$$
$$Q_{ext} = \frac{2}{(kR)^2} \sum_{l=1}^{\infty} (2l+1) \operatorname{Re}(a_l + b_l), \qquad (1)$$
$$Q_{abs} = Q_{ext} - Q_{scs},$$

где электрические $a_l$ и магнитные $b_l$ амплитуды рассеяния для немагнитного материала с диэлектрической проницаемостью $\varepsilon = n^2$ ($n$ есть показатель преломления материала наночастицы) могут быть выражены в виде:

$$a_l = \frac{R_l^{(a)}}{R_l^{(a)} + iT_l^{(a)}}, \quad b_l = \frac{R_l^{(b)}}{R_l^{(b)} + iT_l^{(b)}}, \qquad (2)$$

и функции $R_l$ и $T_l$ могут быть выписаны в следующем виде:

$$R_l^{(a)} = n\psi_l'(kR)\psi_l(nkR) - \psi_l(kR)\psi_l'(nkR),$$
$$T_l^{(a)} = n\chi_l'(kR)\psi_l(nkR) - \chi_l(kR)\psi_l'(nkR), \qquad (3)$$

$$R_l^{(b)} = n\psi_l'(nkR)\psi_l(kR) - \psi_l(nkR)\psi_l'(kR),$$
$$T_l^{(b)} = n\chi_l(kR)\psi_l'(nkR) - \chi_l'(kR)\psi_l(nkR). \qquad (4)$$

Здесь, $\psi_l(x) = \sqrt{\frac{\pi x}{2}} J_{l+1/2}(x)$, $\chi_l(x) = \sqrt{\frac{\pi x}{2}} N_{l+1/2}(x)$, $J_{l+1/2}(x)$ и $N_{l+1/2}(x)$ есть функции Бесселя и Неймана, а штрих означает взятие производной по соответствующему аргументу.

Используя выражения (1)-(4) проведем сравнительный анализ сечения рассеяния и сечения поглощения плазмонной и высокоиндексной диэлектрической наночастицы. В качестве материала плазмонной наночастицы мы выбрали серебро (Ag), имеющее в оптической области спектра наименьшие потери [81] среди других благородных металлов. В качестве материала диэлектрической наночастицы мы используем кристаллический кремний (c-Si), так же проявляющий наилучшие характеристики в видимой области спектра [60,82] среди доступных диэлектрических материалов. Результаты вычислений приведены на Рис. 1. (а),(в) и демонстрируют нормированное полное сечение рассеяния ($Q_{sct}$) серебряной наночастицы (а) и кремниевой наночастицы (в) в зависимости от их радиуса (R) и длины волны. Видно, что серебряная наночастица имеет резонансы исключительно электрического типа, включая электрический дипольный (ed), электрический квадрупольный (eq) и электрический октупольный (eo) резонансы, поведение которых с ростом радиуса схематически показано белыми штриховыми линиями. Напротив, спектр резонансных мод диэлектрической наночастицы значительно шире и включает в себя магнитный дипольный момент (md), который в этом случае является фундаментальным, а так же магнитный квадрупольный (mq) момент и моменты более высокого порядка. Рис.1 (б),(г) демонстрируют результаты вычисления нормированных сечений поглощения серебряной наночастицы (б) и кремниевой наночастицы (г) в зависимости от их радиуса и длины волны. Сравнительный анализ этих результатов показывает, что кремниевая наночастица имеет значительно меньшие диссипативные потери во всей области изменения параметров. Кроме того, диссипативные потери кремниевой наночастицы быстро падают с увеличением длины волны, что объясняется низкими потерями этого материала вдалеке от основной полосы поглощения. Заметим, что золотые Au наночастицы обладают еще большими диссипативными потерями.

Таким образом, плазмонные и диэлектрические наноструктуры обладают как достоинствами, так и недостатками. Недостатком плазмонных наноструктур являются высокие диссипативные потери и как следствие низкие пороги разрушения. Недостатком диэлектрических наноструктур с резонансами Ми являются их низкий коэффициент локального усиления поля. Кроме того, в диэлектрических наноструктурах поле усиливается преимущественно внутри частицы, затрудняя использование таких структур для рамановской спектроскопии и биологических датчиков. Поэтому в последнее время ведутся разработки гибридных металлодиэлектрических наноструктур, объединяющих в себе преимущества металлических и диэлектрических наноантенн (Рис. 2(а)), способных выполнять эффективное преобразование падающего поля в ближнее и обратно, и управлять магнитной компонентой света. В настоящее время такие гибридные наноструктуры находят применение для линейного управления излучением в дальней зоне (Рис. 2(б)), для управления скоростью распада возбужденного состояния квантового источника (Рис. 2(в)), нелинейного управления рассеянием света при помощи света (Рис. 2(г)), а также для генерации новых оптических гармоник (Рис. 2(д)). Перечисленные области применения гибридных наноструктур подробно обсуждаются в следующих главах данной работы.

Из описанного выше следует, что гибридные наночастицы обладают значительно более богатым модовым составом. Исследование модового состава гибридных наноструктур, в том числе нетривиальной формы, проводится с использованием мультипольного разложения численно рассчитанных ближних и дальних электромагнитных полей различными методами [83,84]. Одним из них является метод, основанный на разложении рассеянного в дальней зоне электрического поля $\mathbf{E}_{sca}$, спроецированного на сферическую поверхность, окружающую гибридную наноструктуру, в базисе сферических функций $Y_{lm}$ [85,86]. Мультипольные коэффициенты $a_E$ и $a_M$ в этом случае вычисляются согласно следующим выражениям:

$$a_E(l, m) = \frac{(-i)^{l+1} kR}{h_l^{(1)}(kR) E_0 \sqrt{\pi(2l+1)(l+1)l}} \int_0^{2\pi} \int_0^\pi Y_{lm}^*(\theta, \varphi) \mathbf{r} \mathbf{E}_{sca}(\mathbf{r}) \sin\theta \, d\theta d\varphi \qquad (5)$$

$$a_M(l, m) = \frac{(-i)^l kR}{h_l^{(1)}(kR) E_0 \sqrt{\pi(2l+1)}} \int_0^{2\pi} \int_0^\pi \mathbf{X}_{lm}^*(\theta, \varphi) \mathbf{E}_{sca}(\mathbf{r}) \sin\theta \, d\theta d\varphi \qquad (6)$$

где R – радиус наночастицы, k – волновое число, $h_l^{(1)}$ – функция Ганкеля, $E_0$ – амплитуда падающей волны, $Y_{lm}$ и $\mathbf{X}_{lm}$ – скалярная и векторная сферические функции, $\mathbf{r}$ – радиус вектор, $\theta$ и $\varphi$ – сферические координаты.

Типичным примером гибридной наночастицы является совмещенная система диэлектрической и плазмонной наночастиц различной формы. Например, в работе [85] была исследована гибридная наночастица состоящая из кремниевого наноконуса и золотого нанодиска, помещенного на меньшее основание наноконуса, рис. 3. Было показано, что лазерный нагрев золотой наночастицы позволяет изменять форму золотой наночастицы контролируемым образом с нанодиска (рис. 3(б)) на наночашку (рис. 3(в)) и наносферу (рис. 3(г)). Примеры мультипольных разложений ближних электромагнитных полей кремниевого наноконуса и гибридной наночастицы показаны на рис. 3(а) и рис. 3(б-г), соответственно. В данном случае мультипольные разложения помогают оценить вклады различных мультиполей (md, ed, eq) в спектр рассеяния, а также определить тип каждого из резонансов. Анализ оптических резонансов гибридных наноантенн показывает, что такие наноантенны обладают как электрическим и магнитным резонансами диэлектрической части, так и локализованным поверхностным плазмонным резонансом металлической части, спектральное положение которого может быть изменено посредством лазерной абляции. Причем, при перекрытии магнитной дипольной моды диэлектрической наночастицы и плазмонной моды металлической наночастицы наблюдается их гибридизация, что приводит к возникновению в системе дополнительного электрического квадрупольного отклика. Распределения ближнего электрического поля на соответствующих резонансах системы приведены на рис. 3(А-З). Видно, что в ходе модификации золотой наночастицы изменяются как спектральные свойства рассеянного света так и распределения ближнего поля.

В следующих главах данной работы мы систематически обсуждаем различные области применения гибридных наноструктур. В конце работы мы делаем выводы по проведенному обзорному исследованию и прогнозируем дальнейшие пути развития данной захватывающей и многообещающей области современной науки о взаимодействии света и материи на наноуровне.

## 3. Гибридные наноструктуры для управления характеристиками излучения

Гибридные металло-диэлектрические наноструктуры с диэлектрической и металлической компонентами открывают новые возможности для проектирования устройств с заданными рассеивающими свойствами и диаграммами направленности излучения. Уникальная модовая структура, появляющаяся в гибридных наноантеннах в результате взаимодействия плазмонного резонанса и резонансов Ми диэлектрической наночастицы, формирует в спектре каналы рассеяния, каждый из которых характеризуется своей диаграммой направленности, что позволяет в конечном итоге пространственно разделять приходящие оптические сигналы на различных частотах, а также концентрировать и направлять излучение квантового эмиттера в заданную область в дальней зоне. Такие свойства гибридных наноантенн делают возможной реализацию на их основе источника Гюйгенса [87], суперрассеивателя падающего излучения [88] и оптических разветвителей [89].

Наноантенны и наноструктуры, работающие в качестве приемников и разветвителей оптического сигнала, играют ключевую роль в полностью оптической обработке информации [2,90–92]. Для подобных устройств, с одной стороны, требуется обеспечить достаточно мощное усиление сигнала и его фокусировку и, с другой стороны, пространственную и частотную селекцию сигналов. Объединить в себе эти свойства способны гибридные наноантенны: металлическая часть гибридной наноантенны, благодаря возбуждению поверхностного плазмона, локализует энергию оптического сигнала в субволновых масштабах [93,94], в то время как диэлектрическая часть направленно рассеивает падающие световые волны [95].

На рис. 4(а) изображена гибридная металлодиэлектрическая наноантенна, состоящая из золотого наностержня и кремниевой наносферы, расположенной на его вершине [96]. Золотой наностержень имеет две основные моды: продольную моду и локализованную в зазоре между стержнем и сферой плазмонную моду, природа которых является полностью электрической, в отличие от кремниевой наносферы, поддерживающей как электрическую, так и магнитную моды. Ввиду того, что расстояние между этими наночастицами много меньше длины волны в видимом диапазоне, наведенные в них электрические и магнитные диполи взаимодействуют между собой. Это приводит к гибридизации локализованной в зазоре плазмонной моды наностержня и магнитной и электрической мод наносферы и сдвигу гибридных мод в низкочастотную область спектра (рис 4(б)). В то же время продольная плазмонная мода не претерпевает изменений, поскольку ее возбуждение зависит только от аспектного соотношения наностержня и показателя преломления окружающей среды. Таким образом, в гибридной наноструктуре появляются четыре канала рассеяния, каждый из которых соответствует частоте возбуждения моды. Однако, объектом дальнейших исследований является обеспечение высокой добротности резонансов данной системы, что позволит улучшить селекцию спектрально уплотненных оптических сигналов и сделает возможным его использование в фотонных интегральных схемах.

Димерная гибридная наноструктура с пространственным разделением оптических сигналов в ближнем инфракрасном диапазоне длин волн показана на Рис. 4(в) [89]. Эта наноструктура включает в себя два гибридных диска с кремниевым основанием, серебряной вершиной и промежуточным слоем из оксида алюминия ($Al_2O_3$). В отличие от системы наностержень-сфера рассмотренной выше, резонансы которой были преимущественно дипольного типа, эти диски поддерживают в рабочем диапазоне длин волн так называемые моды шепчущей галереи, сосредоточенные, в основном, в слое из $Al_2O_3$ и усиленные благодаря локализации поля серебряными и кремниевыми частями наноантенны. Из-за сильного взаимодействия между дисками в димерной наноструктуре, мода шепчущей галереи разделяется на две пары. Одна пара соответствует симметричному распределению максимумов и минимумов мод, возбужденных в дисках, относительно плоскости симметрии димера (см. вставки на рис. 4(г)). Вторая пара соответствует антисимметричному распределению поля. Интерференция мод на различных длинах волн дает различные диаграммы направленности излучения в дальней зоне, что может быть использовано для пространственного разделения сигналов (Рис. 4(г)). Благодаря высокой добротности моды шепчущей галереи, данная наноструктура характеризуется высокой частотной избирательностью в узком диапазоне длин волн, что делает подобные гибридные димеры решением для разделения сигналов в системах со спектральным уплотнением DWDM (Dense Wavelength Division Multiplexing).

Для некоторых практических приложений, например, для оптической обработки информации [31,54,97], направленного запуска волноводных мод [98] и фотовольтаики [99] особенно важным является однонаправленное рассеяние падающего излучения, что сложно реализовать в полностью диэлектрических или металлических наноструктурах без использования специально сконструированного отражателя [100–103] или сложной конструкции на подобие антенны Уда Яги. Этот факт объясняется тем, что в стандартных наноструктурах на рабочей длине волны доминирует мультипольный момент одного типа. Поскольку диаграмма направленности мультиполя симметрична, то она имеет как минимум два лепестка, как в случае диполя в E-плоскости, в таких наноструктурах существует как рассеяние вперед, так и назад. Однако, совместив, например, магнитный и электрический дипольные резонансы в гибридной наночастице типа ядро оболочка (Рис. 5(а)), можно добиться их конструктивной интерференции в прямом направлении и деструктивной интерференции в обратном направлении [88,104]. Кроме того, было показано, что в составных наноструктурах, олигомерах из наночастиц ядро оболочка, появляются дополнительные лепестки в диаграмме направленности, что приводит к разветвлению оптического сигнала (Рис. 5(б)).

В работе [105] изучались асимметричные гибридные димерные наноструктуры, включающие золотую и кремниевую сферические наночастицы (Рис. 5(в)), называемые *димер Януса (Juanus nanodimer).* Аналогично вышеописанной наноантенне типа ядро-оболочка, подавление заднего лепестка диаграммы направленности достигается благодаря деструктивной интерференции оптических волн, рассеянных на магнитном дипольном моменте диэлектрической наносферы, и волн, рассеянных на электрическом дипольном моменте металлической наносферы. Из-за относительно малой спектральной ширины магнитной дипольной моды однонаправленное рассеяние имеет узкополосный характер.

Более того, в зазоре димера существует многократно усиленное электрическое поле, что может быть использовано для возбуждения энергетических переходов флуоресцентных молекул и других квантовых источников света. Было продемонстрировано, что однонаправленное рассеяние усиливается при создании цепочек с большим количеством димеров Януса. Это объясняется тем, что дополнительная интерференция волн, рассеянных на соседних димерах, устраняет рассеяние света в боковом и заднем направлениях и усиливает излучение в прямом направлении. Детальное исследование модовой структуры такого димера с помощью обобщенной модели диполь-дипольного взаимодействия, а также его взаимодействие с квантовыми эмиттерами подробно разобрано в работе [106].

Эффективное преобразование свободно-распространяющейся оптической волны в направленную волноводную моду является важной частью полностью оптической передачи информации между двумя логическими элементами. Для решения этой задачи было предложено использовать множество различных конструкций (в том числе и оптических наноантенн), включающих плазмонный волновод [2,5,107,108], способный локализовать свет в субволновых масштабах и передать на микрометровые расстояния с приемлемым затуханием. От наноантенн, осуществляющих направленный запуск поверхностного плазмона, требуется обеспечить синхронизм возбуждающей световой волны и волноводной моды, многократное усиление и концентрацию волны в заданном направлении. На Рис. 6(а) изображена металлодиэлектрическая наноантенна, в основе которой лежит кремниевый нанодиск, находящейся над поверхностью серебряной пленки [109]. При падении на данную наноструктуру оптической волны с TM-поляризацией под скользящими углами в зазоре между диском и пленкой возбуждаются гибридные моды, добротность которых может достигать $10^3$, что выражается в гигантском усилении напряженности электрического поля до значений порядка 60 по сравнению с напряженностью падающей волны (рис. 6(б)). Гибридная мода, в свою очередь, запитывает моду плазмонного волновода, преобразовываясь в поверхностный плазмон-поляритон, распространяющийся преимущественно в одном направлении (рис. 6(в)).

## 4. Гибридные наноструктуры для управления характеристиками ближнего поля

Локализация световой энергии в ближнем поле и связанные с этим эффекты, такие как увеличение эффективности возбуждения квантовых источников [97], усиление фотолюминесценции [110] и эффект Перселла [56], являются одним из важнейших свойств металлических наноантенн. Это обусловлено возбуждением в таких наночастицах поверхностных плазмонов, которые с одной стороны концентрируют падающее электромагнитное поле на поверхности металла, и с другой стороны накапливают и переизлучают энергию квантового эмиттера, находящегося вблизи наноантенны. Общая теория эффекта Перселла, заключающегося в изменении скорости спонтанной эмиссии квантового эмиттера вблизи резонансной наноантенны приведена ниже в этой Главе. В то же время диэлектрические наноантенны, характеризующиеся высокой направленностью излучения и малыми диссипативными потерями, не способны обеспечить такую же степень локализации поля, как металлические. Поэтому в таких областях, как поверхностно усиленная рамановская спектроскопия (SERS) и фотолюминесцентная спектроскопия, а также полностью оптическая обработка информации, начал развиваться подход, объединяющий диэлектрические компоненты, сделанные из полупроводников с высоким показателем преломления и низким коэффициентом экстинкции, и металлические компоненты в единой гибридной наноструктуре. Такой подход позволил решить ряд задач, начиная от управления спонтанной эмиссией квантового источника [111] и создания эффективных преобразователей энергии возбужденного состояния в оптический сигнал [112] для применения в фотонных вычислениях, и заканчивая локальным возбуждением [113] флуоресцентных меток для микроскопии со сверхразрешением и концентрацией оптической энергии в солнечных батареях [113].

**Общая теория эффекта Перселла**. Как отмечалось ранее, эффект Перселла один из ключевых эффектов в квантовой оптике и нанофотонике. Сильный эффект Перселла обычно наблюдается при помещении источника в микрорезонаторы [114–116], горячие пятна резонансных наноантенн [9,53,117–125], фотонные кристаллы [126–132] или метаматериалы [49,122,133–137]. В оптической области частот измерение скорости спонтанного излучения обычно проводится с помощью записи временного сигнала фотолюминесценции от источника, возбуждённого в импульсном режиме [138–140]. В микроволновом и ТГц диапазонах, наоборот, эффект Перселла можно наблюдать как увеличение излучаемой мощности антенной в стационарном режиме [48]. С практической точки зрения увеличение скорости спонтанного излучения может быть применено в лазерах, однофотонных источниках, флуоресцентной микроскопии, биологических исследованиях и спектроскопии. В настоящее время в литературе можно найти несколько детальных обзорных работ [48,53,134,141], а так же книг [142] посвященных эффекту Перселла и модификации спонтанной эмиссии, поэтому здесь мы коснемся только основных моментов, необходимых для понимания дальнейшего изложения.

До недавнего времени акцент в изучении эффекта Перселла делался на скорость спонтанного излучения электрических дипольных переходов. Это связано с тем, что обычно у квантовых источников электрические дипольные переходы значительно сильнее, чем магнитные. Это различие обуславливает то, что магнитная проницаемость большинства материалов равна 1 в видимом диапазоне длин волн. Поэтому контролировать магнитный отклик источников гораздо труднее, чем электрический. Однако, существует ряд квантовых источников, таких как редкоземельные ионы, полупроводниковые квантовые точки, у которых магнитные переходы сравнимы или даже больше, чем электрические [143–147]. В работе [53] был дан широкий обзор на возможные источники с доминирующими магнитными переходами.

Простейшая модель квантовых источников, отображающая большинство свойств реальных квантовых оптических излучателей – это двухуровневая система с основным состоянием |g> и возбуждённым состоянием |e>, разница энергии между которыми равна $\hbar\omega_0$. Переходные процессы характеризуются матричным дипольным элементом $\boldsymbol{d}_{eg} =$

$\langle e|qr|g\rangle$. Основные свойства такой системы заключаются в том, что при релаксации возбужденного состояния в основное, испускается фотон с энергией $\hbar\omega_0$. Для расчета подобных процессов нужно учитывать взаимодействие двухуровневой системы с континуумом состояний свободного пространства. Эти расчеты были выполнены Вайскопфом и Вигнером в работе [148]. Они показали, что возбужденные состояния двухуровневой системы, расположенной в свободном пространстве распадаются со скоростью $\gamma_0$:

$$\gamma_0 = \frac{\omega_0^3}{3\pi\hbar\varepsilon_0 c^3}|\boldsymbol{d}_{eg}|^2 \qquad (7)$$

где, $\varepsilon_0$ – диэлектрическая проницаемость вакуума. В случае, когда двухуровневая система находится в какой-то среде, скорость спонтанного излучения изменяется. Здесь и далее мы используем систему единиц СИ, так как именно она чаще всего используется в этой области. Для того чтобы перейти в систему СГС достаточно произвести замену $\varepsilon_0$ на $1/(4\pi)$. Используя золотое правило Ферми, эта скорость может быть записана следующим образом:

$$\gamma = \frac{\pi\omega_0}{\hbar\varepsilon_0}|\boldsymbol{d}_{eg}|^2 \rho_{\boldsymbol{n}}(\boldsymbol{r}_0, \omega_0), \qquad (8)$$

где, $\rho_{\boldsymbol{n}}(\boldsymbol{r}_0, \omega_0)$ – локальная плотность состояний электромагнитного поля в точке расположения системы $\mathbf{r}_0$:

$$\rho_{\boldsymbol{n}}(\boldsymbol{r}_0, \omega_0) = \sum_{\boldsymbol{k}}[\boldsymbol{n}\cdot\boldsymbol{e}_k(\boldsymbol{r}_0)\otimes\boldsymbol{e}_k^*(\boldsymbol{r}_0)\cdot\boldsymbol{n}]\delta(\omega_k - \omega_0), \qquad (9)$$

где, суммирование производится по всем собственным модам системы $\boldsymbol{e}_k$ с собственными частотами $\omega_k$. Собственные состояния $\boldsymbol{e}_k$ – это решение однородного волнового уравнения, нормированное условием $\int_V \varepsilon(\boldsymbol{r})\boldsymbol{e}_i(\boldsymbol{r})\cdot\boldsymbol{e}_j(\boldsymbol{r})d^3r = \delta_{ij}$, где $\varepsilon(\boldsymbol{r})$ – диэлектрическая проницаемость среды. Единичный ветор **n** ориентирован в направлении дипольного момента $\boldsymbol{d}_{eg}$. Меняя локальную плотность состояний, можно значительным образом модифицировать скорость спонтанного излучения.

Выражение (8) дает правильный результат скорости излучения в случае режима слабой связи, когда константа взаимодействия между излучателем и электромагнитными состояниями меньше, чем скорость распада электромагнитных состояний. Такой режим соответствует динамике Маркова, когда система не запоминает эволюцию во времени и затухает экспоненциально. В противоположном случае, сильное взаимодействие между излучателем и электромагнитными состоянии, приводит к случаю не Марковской динамики, и уравнение (8) не может быть использовано для описания скорости излучения. В этом случае источник может не иметь экспоненциальное затухание, а также могут присутствовать осцилляции Раби между состояниями [149–151].

Электромагнитная локальная плотность состояний в формуле (9) может быть рассчитана с использованием диадного тензора функции Грина электрического источника, который может быть разложен по собственным состояниям (модам) системы [152]. Для резонаторов без потерь, диадная функция Грина принимает следующий вид:

$$\boldsymbol{G}(\boldsymbol{r},\boldsymbol{r}',\omega) = \sum_k c^2 \frac{\boldsymbol{e}_k^*(\boldsymbol{r}) \otimes \boldsymbol{e}_k(\boldsymbol{r}')}{\omega_k^2 - \omega_0^2}, \qquad (10)$$

где, $\boldsymbol{G}(\boldsymbol{r},\boldsymbol{r}',\omega)$ – тензорная функция Грина, решение неоднородного волнового уравнения:

$$\boldsymbol{\nabla} \times \boldsymbol{\nabla} \times \boldsymbol{G}(\boldsymbol{r},\boldsymbol{r}',\omega) - \varepsilon(\boldsymbol{r})\frac{\omega_0^2}{c^2}\boldsymbol{G}(\boldsymbol{r},\boldsymbol{r}',\omega) = \boldsymbol{I}\delta(\boldsymbol{r}-\boldsymbol{r}'), \qquad (11)$$

где $\boldsymbol{I}$ – единичный тензор. Собственные состояния, входящие в выражение (9) позволяют представить локальную плотность состояний, как:

$$\rho_{\boldsymbol{n}}(\boldsymbol{r}_0,\omega_0) = \frac{2\omega_0}{\pi c^2}\boldsymbol{n} \cdot \mathrm{Im}\,\boldsymbol{G}(\boldsymbol{r}_0,\boldsymbol{r}_0,\omega_0) \cdot \boldsymbol{n}. \qquad (12)$$

Используя формулу (12), результирующие выражение для скорости спонтанного излучения можно записать в виде:

$$\gamma = \frac{2\omega_0^2}{\hbar\varepsilon_0 c^2}|\boldsymbol{d}_{eg}|^2 \mathrm{Im}\,\boldsymbol{G}(\boldsymbol{r}_0,\boldsymbol{r}_0,\omega_0) \cdot \boldsymbol{n}. \qquad (13)$$

Удобной характеристикой для описания изменения скорости спонтанного излучения является безразмерная величина, которая называется фактор Перселла $F_p$. Данная величина определяется как отношение скорости излучения двухуровневой системы в неоднородной среде (например, в присутствии наноантенны) к скорости в свободном пространстве $\gamma_0$:

$$F_p = \frac{\gamma}{\gamma_0} = \frac{6\pi c}{\omega_0} \boldsymbol{n} \cdot \operatorname{Im} \boldsymbol{G}(\boldsymbol{r}_0, \boldsymbol{r}_0, \omega_0) \cdot \boldsymbol{n}. \tag{14}$$

Стоит отметить, что значение $F_p$ (в режиме слабой связи) не зависит от дипольного момента двухуровневой системы и определяется только электромагнитными свойствами окружающей среды.

Для открытых систем или в случае присутствия потерь выражение (10) не может быть использовано вследствие того, что невозможно определить базис собственных состояний $\boldsymbol{e}_k$. Тем не менее, метод функции Грина, используемый в уравнении (14), даёт верный результат для скорости излучения электрического дипольного перехода у источника, расположенного в открытом резонаторе.

Выражение (14) дает возможность получить скорость излучения квантового источника, зная классическую характеристику - тензорную функцию Грина. Кроме того, это выражение можно использовать для интерпретации эффекта Перселла в классическом случае. Увлечение скорости излучения можно понимать как увеличение работы совершаемой электрическим полем. Работа, совершаемая электрическим полем осциллирующего диполя $\boldsymbol{d}e^{-i\omega_0 t}$, дается следующим выражением:

$$P = \frac{\omega}{2} \operatorname{Im}[\boldsymbol{d}^* \boldsymbol{E}(\boldsymbol{r}_0)] = \mu_0 \frac{\omega^3}{2} |\boldsymbol{d}|^2 \boldsymbol{n} \cdot \operatorname{Im} \boldsymbol{G}(\boldsymbol{r}_0, \boldsymbol{r}_0, \omega_0) \cdot \boldsymbol{n}, \tag{15}$$

где $\mu_0$ – проницаемость вакуума. Деля это значение на мощность, излучаемую тем же источником в свободном пространстве $P_0 = \frac{\omega_0^4}{12\pi\varepsilon_0 c^3}|\boldsymbol{d}|^2$, мы получаем выражение идентичное формуле (14).

Полная скорость распада γ, входящая в уравнение (13) включает в себя два члена: излучательный вклад и безызлучательный. В большинстве практических случаев более интересным представляется увеличение излучаемой мощности источника в дальнюю зону. Излучательный фактор Перселла в этом случае определяется как $F_p^{(r)} = \frac{\gamma_r}{\gamma_0}$, где $\gamma_r$ – излучательная скорость распада, которая может быть посчитана путём интегрирования вектора Пойнтинга по поверхности, охватывающей источник и неоднородное окружение. Для описания части энергии, которая излучается в виде фотонов используется величина, называемая *квантовым выходом*, она определяется следующим образом:

$$\eta_0 = \frac{\gamma_r}{\gamma_r + \gamma_{nr} + \gamma_{int}}, \qquad (16)$$

где $\gamma_{nr}$ – скорость безызлучательного распада, которая описывает электромагнитные потери среды, а $\gamma_{int}$ – внутренняя скорость безызлучательного распада, которая существует даже в отсутствии неоднородного окружения (в свободном пространстве). Квантовый выход описывает часть энергии, излученной изолированной двухуровневой системой. В случае, когда система помещена в какую-то среду, выражение для квантового выхода принимает следующий вид:

$$\eta = \frac{F_p^{(r)}}{\eta_0 F_p + (1 - \eta_0)}. \qquad (17)$$

Выражение (14) для фактора Перселла может быть упрощено в случае, когда главный вклад в тензорную функцию Грина источника дает одна мода. В этом особом случае может быть использовано одномодовое приближение, в котором тензорная функция Грина записывается следующим образом:

$$\boldsymbol{G}(\boldsymbol{r}, \boldsymbol{r}', \omega) \approx c^2 \frac{\boldsymbol{e}_k^*(\boldsymbol{r}) \otimes \boldsymbol{e}_k(\boldsymbol{r}')}{\omega_k^2 - \omega_0^2 - 2i\gamma_k \omega}, \qquad (18)$$

где, $\gamma_k$ – это скорость накачки у соответствующей собственной моды $\boldsymbol{e}_k(\boldsymbol{r})$. Учитывая, что частота источника совпадает с частотой моды резонатора, а источник сонаправлен с создаваемым полем, мы приходим к известной формуле для одномодового фактора Перселла:

$$F_p = \frac{3}{4\pi^2} \lambda^3 \frac{Q}{V},  \qquad (19)$$

где $Q = \omega_k/2\gamma_k$ -добротность моды, $\lambda$ - длина волны в свободном пространстве и V – эффективный объем моды, который в приближении малых потерь определфется как[153,154]

$$V = \frac{\int \varepsilon(r)|E(r)|^2 \, dV}{\max\left(\varepsilon(r)|E(r)|^2\right)}. \qquad (20)$$

Обычно эффективный объем моды много меньше физического объема резонатора. Например, плазмонные нанорезонаторы способны поддерживать моды с крайне малыми значениями объема вплоть до $\lambda^3/10^4$ [155]. Заметим, что выражение (20) не всегда применимо к реальным оптическим резонаторам с потерями, особенно к плазмонным наночастицам, поскольку последние имеют высокие диссипативные потери [156]. По этой причине одномодовое приближение не всегда даёт точный результат, в частности, для источников, расположенных близко к плазмонным наночастицам. Обсуждению этого вопроса было посвящено несколько работ [52,155–160]. В частности, детальный анализ эффективного объема моды плазмонных нанорезонаторов представленый в работе [155], показал необходимость введения так называемого комплексного объема моды.

Отметим, что резонансные состояния являются ненормированными в традиционном смысле, что накладывает на формулу (20) некоторые ограничения. Это связанно с тем, что поле собственных электромагнитных резонансных состояний открытых резонаторов экспоненциально расходится при удалении от системы. Проблема нормировки резонансных состояний исследовалась и в итоге была решена многими авторами, начиная с работ Я. Б. Зельдовича[161]. Применительно к электродинамике это было сделано относительно недавно в работах [162,163].

**Гибридные наноструктуры для управления характеристиками ближнего поля**. Простейшая реализация концепции гибридной наноантенны, включающей золотую наносферу и планарную диэлектрическую антенну представлена на Рис. 7(а). В такой конфигурации квантовый эмиттер помещен в поверхностном слое диэлектрической антенны, которая выполняет две функции. Во-первых, она увеличивает скорость спонтанной эмиссии эмиттера в n раз (n - показатель преломления диэлектрика) [111,164]. Во-вторых, диэлектрическая антенна локализует поле поверхностного плазмона металлической наночастицы в области зазора между ними, что создает дополнительное увеличение фактора Перселла. Данная система служит в качестве излучающей наноантенны, при этом металлическая наночастица выполняет функцию директора стандартной радиоволновой антенны, а планарная диэлектрическая часть – роль направляющего элемента (рефлектора).

Гибридная наноантенна, изображенная на Рис. 7(б), обладает более совершенной конструкцией, что позволяет ей не только увеличивать фактор Перселла, но и концентрировать и передавать излучение в заданном направлении с помощью диэлектрической (Si) наночастицы [87,165,166]. В качестве оптических элементов антенны присутствуют кремниевый диск и золотой наностержень, расположенный над диском параллельно его поверхности на оптимальном расстоянии. Для фиксации наностержня на заданной высоте вся структура помещена в стеклянную оболочку. В такой наноструктуре металлический наностержень играет роль усилителя оптического сигнала и ускоряет спонтанную эмиссию квантового источника (показан красной стрелкой), преобразуя энергию его возбужденного состояния в поверхностный плазмон. Плазмон, затем, переизлучает оптические волны, запитывая электрическую квадрупольную моду кремниевого нанодиска с узконаправленной диаграммой мощности излучения. В результате, выходной сигнал получается многократно усиленным и сконцентрированным в одном направлении. Для изготовления данной гибридной наноантенны применяется двухступенчатая электронно-лучевая литография, при которой сначала на поверхности подложки изготавливается кремниевый диск, после чего на структуру осаждается слой стекла, и затем методом spin-coating на стекле создается металлический наностержень [165].

Микрорезонаторы наряду с металлическими наноантеннами часто используются для усиления взаимодействия между световой волной и квантовым эмиттером из-за сильного эффекта Перселла в таких структурах. Диэлектрические дисковые резонаторы обеспечивают высокие значения фактора Перселла благодаря высокой добротности Q моды шепчущей галереи, которая в них возбуждается, но, в то же время, и высокие значения объема моды V. Для достижения большего фактора Перселла дисковый микрорезонатор и металлическая наноантенна, обладающая, как уже было отмечено, низким V, были объединены в одну гибридную систему (Рис. 7(в)) [112]. Принцип работы и структура гибридного устройства сходна с рассмотренной ранее: золотой наностержень эллипсоидной формы размещен над микроразмерным диэлектрическим диском из нитрида кремния ($Si_3N_4$) с показателем преломления n=1.997, квантовый эмиттер находится возле одного из концов наностержня. Взаимодействие локализованной плазмонной моды металлической наноантенны и высокодобротной моды шепчущей галереи приводит к усилению эмиссии более чем в 1000 раз.

В работе [167] теоретически предложена концепция гибридной металл-диэлектрической наноантенны «галстук-бабочка» (Рис. 8(а)). Такая наноантенна представляет собой традиционную плазмонную наноантенну в форме галстук-бабочки, наконечники которой выполнены из алмаза, содержащего азото-замещенную вакансию (так называемый NV-центр, от англ. nitrogen vacancy center [168,169]). NV-центры в алмазе являются перспективным кандидатом на роль квантовых источников, благодаря простой структуре их энергетических уровней и возможности управления электронными спинами центров посредством микроволнового излучения [170] и, вследствие этого, модуляции спектров поглощения и люминесценции. Авторы данной работы показали, что в такой наноантенне объем моды весьма мал, и электрическое поле сконцентрировано в центре галстука бабочки (Рис. 8б), где сосредоточены NV-центры. Это увеличивает фактор

Перселла до значений порядка 110, а также повышает эффективность сбора излученных фотонов в 1.77 раза.

Среди различных оптических резонансных явлений в ансамблях сильновзаимодействующих наноструктур особое внимание привлекает *резонанс Фано*, впервые предложенный Уго Фано в 1961 году [171] для описания ярко выраженной асимметрии профиля спектров поглощения благородных газов, но затем обобщенный на другие физические системы, включая оптические наноструктуры [79,172–179]. Было показанно, в частности, что в оптике это явление приводит к узким спектральным особенностям в спектрах рассеяния, отражения или поглощения, имеющим принципиальное значение для химических и биологических сенсоров. Как правило, резонанс Фано в наноструктурах возникает в результате интерференции двух мод: широкой нерезонансной моды и узкой резонансной [66,171,173,175,179–185]. При этом наблюдаются противофазные колебания мод, возбужденных в наноструктурах, что выражается в минимальном переизлучении оптической энергии в дальнюю зону и появлении, так называемого, провала Фано в спектре. Рис. 8(в) иллюстрирует гибридную наноантенну, состоящую из кремниевой сферической наночастицы помещенной внутрь золотого кольца. Вся структура имеет субволновые размеры. Структурной особенностью этой наноантенны является наличие зазора между внутренними стенками кольца и сферы. Из-за низкодобротных плазмонных резонансов золотое кольцо обладает оптическим откликом в широком диапазоне длин волн. В то же время, магнитный дипольный резонанс Ми кремниевой сферы имеет относительно узкую ширину. Сильное взаимодействие между магнитной дипольной и плазмонной модами приводит к появлению магнитно-электрического резонанса Фано, сопровождающегося мощным усилением как электрического, так и магнитного полей в свободном пространстве зазора между кольцом и сферой (Рис. 8(г)). Такое усиление может быть использовано для увеличения мощности сигнала фотолюминесценции квантовых эмиттеров с электрическими или магнитными переходами [53].

Для приложений в оптоэлектронике, в частности, в фотодетектировании, полностью оптической модуляции и фотовольтаике часто требуется создать высокое поглощение в приповерхностной зоне полупроводниковой подложки. Наноструктурирование поверхности и использование полупроводниковых нановключений и наноантенн не всегда способно обеспечить достаточный уровень поглощения оптического излучения из-за низкой локализации поля в объеме полупроводника. Поверхностные плазмонные резонансы, возникающие в металлических наноструктурах, способны решить данную проблему, поскольку, как многократно отмечалось ранее, они на порядок усиливают оптическое поле вблизи металлической поверхности. Применение металлических покрытий наряду с полупроводниковыми одномерными решетками (Рис. 9(а)) и наноантеннами (Рис. 9(б)) положило начало в этой области новому классу метаповерхностей из гибридных металлодиэлектрических наноструктур, полностью поглощающих падающее на них оптическое излучение (так называемые *идеальные поглотители*, от англ. perfect absorbers) [186,187]. Активные полупроводниковые компоненты, имеющие характерные для диэлектрических наночастиц резонансы Ми (электрические дипольные, в данном случае), поглощают световую энергию, в результате чего в полупроводнике возникает высокая концентрация электронно-дырочных пар. В свою очередь металлические покрытия за счет плазмонных резонансов дипольного типа обеспечивают локализацию и усиление поля в полупроводниковых компонентах (Рис. 9(в,г)), а также, вследствие нерадиационных потерь в металле, инжектируют «горячие» электроны через границу раздела металл-полупроводник, в зону проводимости полупроводника, вызывая тем самым дополнительный фототок, в том числе и при энергиях фотонов меньше ширины запрещенной зоны полупроводника, что особенно важно при создании широкополосных фотодетекторов.

Обе наноструктуры, изображенные на Рис. 9(а,б), имеют свои преимущества и недостатки. Одномерная решетка способна поглощать до 95% падающего на нее излучения в ближнем инфракрасном диапазоне, однако имеет сильную чувствительность к поляризации излучения. В свою очередь двумерная решетка, иначе метаповерхность из гибридных наноантенн, характеризуется меньшими значениями коэффициента поглощения (80-90%), являясь при этом поляризационно-независимой. Фабрикация данных наноструктур происходит в два этапа. Сначала изготавливаются полупроводниковые решетки путем электронно-лучевой литографии и реактивного ионного травления. Затем на полупроводник осаждается 15-нм слой золота поверх 1-нм слоя титана для лучшей адгезии.

Рассмотренные выше наноструктурированные металл-полупроводниковые поверхности являются перспективным кандидатом на роль покрытий для солнечных батарей будущего благодаря высокому поглощению, ведущему к росту *внутреннего квантового выхода*. Тем не менее, достижение максимальной эффективности конверсии оптической энергии в фототок в таких наноструктурах до значений коэффициента поглощения осложнено неэффективным сбором фотоиндуцированных носителей заряда электродами, в результате чего падает общая эффективность солнечного элемента. В работе [188] было предложено возможное решение этой проблемы путем замены поверхностных фотоэлектрических элементов гибридными наностержнями типа ядро-оболочка, основу которых составляют металлические (например, серебряные) электроды, покрытые слоем полупроводника (Рис. 10(а)). Из-за малой толщины полупроводниковой оболочки и наличия двух границ раздела: полупроводник-воздух и металл-полупроводник, поляризационно-зависимые резонансы Ми, характерные для диэлектрических нитей и стержней, вырождаются в поляризационно-независимые резонансы типа Фабри-Перо, которые, как видно из распределений полей на Рис. 10(б,в), можно интерпретировать как магнитный ($TE_{01}$) и электрический ($TE_{11}$) дипольные резонансы. Возможность гибкой перестройки данных резонансов позволяет обеспечить в широком диапазоне длин волн видимого диапазона высокую эффективность поглощения до значений 1.9.

## 5. Нелинейные и перестраиваемые гибридные наноантенны

Генерация высших гармоник имеет потенциал широкого применения в микроскопии биологических молекул [189,190] и визуализации внутриклеточных процессов [191,192], так как позволяет осуществлять накачку фотолюминесцентных красителей на частотах ближнего инфракрасного диапазона в окне прозрачности клеточных стенок. Для подобных приложений необходимо разместить наночастицы из материалов, эффективно преобразующих оптический сигнал основной частоты в сигнал на удвоенной или утроенной частоте, внутри или вблизи исследуемого объекта. К материалам, способным генерировать вторую гармонику относятся нелинейные диэлектрические кристаллы с нецентросимметричной кристаллической решеткой ($LiNbO_3$, $ZnS$ и др). Поскольку коэффициент преобразования зависит от длины оптического пути внутри кристалла, наночастицы не способны к эффективной генерации из-за их малых размеров. Это существенно уменьшает уровень сигнала и, как следствие, вероятность его взаимодействия с молекулами красителя. Поэтому для увеличения эффективности преобразования развиваются новые подходы, одним из которых является размещение вблизи нелинейных частиц резонансных наноантенн или встраивание наноструктурных элементов в саму частицу.

На Рис. 11(а) показана гибридная металлодиэлектрическая наноантенна, состоящая из ортогонально расположенных золотой дипольной антенны и диэлектрической ($ZnS$) дипольной антенны, а так же наночастицы из $ZnS$, ответственных за конверсию оптического сигнала во вторую гармонику [193]. Изготовление такой наноантенны происходит в два этапа. Сначала с помощью электронно-лучевой литографии создаются золотые диполи и специальные позиционные метки. В течение второго этапа эти метки детектируются, и электронно-лучевая литография диполей из $ZnS$ производится в области их позиционирования. Золотой диполь локализует внешнее световое излучение в зазоре, причем величина усиления поля максимальна, когда направление вектора напряженности электрического поля совпадает с осью симметрии металлического диполя. Диэлектрическая антенна, имеющая резонанс на частоте второй гармоники, предназначена для увеличения скорости спонтанной эмиссии диполей, колеблющихся с частотой второй гармоники. Направление поляризации выходного излучения при этом совпадает с осью дипольной диэлектрической антенны. Эти два фактора – усиление поля на основной частоте и увеличение фактора Перселла на удвоенной частоте приводят к росту сигнала второй гармоники в 500 000 раз по сравнению с массивом наночастиц из ZnS.

В статье [194] авторы демонстрируют усиленную генерацию второй гармоники в гибридной металл-диэлектрической димерной наноантенне типа димер Януса (Рис. 11(б)). Совместно с традиционной для таких наноантенн плазмонной сферической наночастицей с размерами порядка 70 нм в димере используется наночастица из титаната бария ($BaTiO_3$) диаметром 100 нм. Для изготовления этой наноантенны применялся метод последовательной капиллярной агрегации Au и $BaTiO_3$ наночастиц в полости, сделанные в полимерном резисте электронно-лучевой литографией. $BaTiO_3$ имеет нецентросимметричную кристаллическую структуру и служит для генерации сигнала на удвоенной частоте, в то время как пламонная наночастица, выполненная из Au усиливает эффект Парселла в системе. В результате, фактор усиления генерации второй гармонике в гибридном димере по сравнению с одиночными наночастицами из $BaTiO_3$ составляет 15. Несмотря на его небольшое значение по сравнению с 500 000 усилением в предыдущей наноантенне, такой димер отличается относительной простотой изготовления и возможностью использования коммерческих Au и $BaTiO_3$ наночастиц.

В отличие от второй гармоники, третья гармоника может быть получена из материалов с центральной симметрией кристаллической решетки. Одним из таких материалов является оксид меди ($Cu_2O$) – диэлектрик с высоким показателем преломления. Однако, как и в случае второй гармоники, для эффективной конверсии в третью гармонику необходимо, чтобы частицы были оптически резонансными в инфракрасном диапазоне, что делает их размеры слишком большими, а сами частицы неприемлемыми, например, для локализации одиночных внутриклеточных процессов. Для преодоления этих недостатков была предложена гибридная наноструктура типа ядро-оболочка, состоящая из золотого ядра в форме стержня и оболочки из $Cu_2O$ (Рис. 11(в)) [195]. Настроенный на плазмонный резонанс в оптическом диапазоне наностержень обеспечивает сильную локализацию электрического поля в ближней зоне. Высокоиндексная оболочка из $Cu_2O$ в такой конструкции смещает плазмонный резонанс в длинноволновую область и поглощает энергию локализованного плазмона, излучая сигнал третьей гармоники.

Аналогичный принцип работы имеет гибридная наноантенна, изображенная на Рис. 11(г). В данной антенне совмещены классическая металлическая дипольная наноантенна, более чем в 30 раз усиливающая электрическое поле в зазоре, и нелинейная наночастица из оксида индия-олова (ITO) с диэлектрической проницаемостью $\varepsilon = 2.9$ на рабочей длине волны, ответственная за конверсию сигнала на основной частоте в сигнал утроенной частоты [196]. Изготовление такой наноантенны происходит в несколько этапов: вначале, традиционной электронно-лучевой литографией на стекле изготавливается металлическая часть наноантенны, после чего покрывается слоем резиста, вновь облучаемого электронными лучами в области зазора диполя. В результате последующего травления в зазорах металлической дипольной антенны образуются отверстия. Капля концентрированного раствора монодисперсных нанокристаллов ITO, связанных с гексаном, осаждается на резист и прокатывается через поверхность образца срезанной частью полидиметилсилоксана (PDMS). Частицы вытягиваются вместе с мениском медленно испаряющегося раствора гексана и осаждаются в прорезанные отверстия. Полученная наноструктура способна усилить нелинейный отклик частицы в $10^6$ раз.

В работе [197] исследовалась генерация третьей гармоники в гибридной металлодиэлектрической наноструктуре, включающей золотое кольцо и кремниевый диск, помещенный в его центр (Рис. 11(д)). Источником оптического сигнала на утроенной частоте в такой конструкции является наночастица кремния. Усиление генерации третьей гармоники в системе кольцо-диск обусловлено сильной локализацией ближних электромагнитных полей в объеме кремниевой частицы и на поверхности золотого диска, а также подавлением излучения в дальнюю зону на основной частоте. Эти эффекты достигаются благодаря возбуждению в наноструктуре так называемого *анаполя* [36,198–205]. Анаполь – это конфигурация двух моментов, тороидального и электрического дипольного моментов, при которой они колеблются в противофазе друг относительно друга. Поскольку диаграммы направленности тороидального и электрического дипольного моментов, при рассмотрении их независимо друг от друга, практически идентичны, волны, излучаемые ими в составе анаполя, гасят друг друга, и оптический сигнал в дальней зоне падает. В то же время, электромагнитная энергия локализуется в ближней зоне анапольной структуры. В гибридной наноантенне, изображенной на рис. 11(д), тороидальный момент возникает в золотом диске на основной частоте, в то время как электрический дипольный момент появляется в кремниевом диске. Гашение вторичных волн на этой частоте и сильная локализация ближнего поля обеспечивают коэффициент конверсии в третью гармонику до 0.007%. Гибридная наноструктура была изготовлена с помощью двухэтапной электронно-лучевой литографии, сопряженной с реактивным ионным травлением. В течение первого этапа, используя электронно-лучевую литографию PMMA-резиста с последующим осаждением хрома и процедурой lift-off, была сделана хромовая маска для последующего ионного травления кремниевых дисков. На втором этапе было выборочно произведено осаждение резиста вблизи дисков, помеченных крестообразными метками. После электронно-лучевой литографии и травления резиста был нанесен слой золота и в результате процедуры lift-off получены золотые кольца [195].

Нелинейные гибридные металлодиэлектрические наноструктуры являются одним из перспективных кандидатов на роль оптических переключателей и логических элементов для полностью оптических интегральных микросхем благодаря возможности управления материальными параметрами таких наноструктур (диэлектрическая проницаемость, показатель преломления) модуляцией интенсивности падающего излучения [206–209]. Поскольку модовая структура наноантенны имеет прямую зависимость от диэлектрической проницаемости ее элементов, изменение интенсивности излучения приводят к сдвигу оптических резонансов наноантенны и изменению диаграммы направленности.

В работе [210] теоретически описана нелинейная димерная наноантенна, состоящая из серебряной (Ag) и кремниевой (Si) наносфер. Нелинейность третьего порядка в серебряной наночастице оказывается достаточно велика, чтобы влиять на ее диэлектрическую проницаемость. На рис. 12(а) схематично изображена такая гибридная наноантенна. Волновой вектор падающего излучения направлен по оси симметрии димера. Размер серебряной наносферы соответствует возбуждению в ней поверхностного плазмонного резонанса на частоте фотонов с энергией 3.14 эВ при низких интенсивностях оптического излучения. В нерезонансном режиме направление наведенного электрического дипольного момента в серебряной наночастице противоположно направлению вектора электрической напряженности падающего поля и наведенному электрическому моменту в кремниевой наночастице. Это ведет к деструктивной интерференции волн, рассеянных электрическими диполями двух частиц и ориентации диаграммы направленности в обратное направление по отношению к волновому вектору падающей волны (Рис. 12(б)). В режиме плазмонного резонанса направление всех электрических дипольных моментов совпадает, рассеянные волны усиливают друг друга, и характер диаграммы направленности становится всенаправленным. Так как спектральное положение плазмонного резонанса определяется в том числе диэлектрической проницаемостью материала наночастицы, ее изменение приводит к сдвигу дипольного резонанса и, в результате, переходу системы из резонансного режима в нерезонансный и обратно. В конечном итоге, это выразится в переключении диаграммы направленности и направления распространения рассеянного света. Причем, скорость переключения составляет 40 фс при относительно низкой интенсивности 6 МВт/см$^{-2}$, что может быть использовано для сверхбыстрых фотонных устройств и цепей. На практике гибридные димеры могут быть изготовлены с помощью недавно предложенной комбинации изготовления сверху вниз и шаблонной самосборки [211], что позволяет точно и контролируемо вертикально и горизонтально позиционировать плазмонные элементы относительно диэлектрических сфер. Кроме того, такие системы могут быть изготовлены путем атомно-силовой наноманипуляции.

Основу нелинейной гибридной наноантенны типа Уда- Яги, представленной на Рис. 12(с), составляет цепочка из наночастиц типа ядро-оболочка (Ag/Ge) [6]. Как было описано ранее, такие наночастицы поддерживают на рабочей длине волны как магнитную дипольную моду диэлектрической оболочки, так и электрическую дипольную моду металлического ядра. Это делает гибридные наночастицы источником однонаправленных оптических волн. В то же время, демонстрируемые в работе [8,68,212] цепочки из полностью диэлектрических сфер, настроенных на магнитные резонансы, имеют сверхнаправленный режим, связанный с возбуждением темной моды и возникновением *сингулярности Ван-Хова*. Цепочка из гибридных наночастиц объединяет в себе свойства сверхнаправленности отдельных ее составляющих, что многократно увеличивает величину направленности такой антенны. При этом если поместить дипольный источник электромагнитных волн в центр такой цепочки, то диаграмма мощности излучения этого источника будет иметь два противонаправленных узких лепестка. Известно, что под действием интенсивного оптического излучения с энергией фотонов, превосходящих ширину запрещенной зоны полупроводника, в нем происходит генерация электронно-дырочной плазмы, что в значительной степени повышает проводимость и уменьшает реальную часть диэлектрической проницаемости на несколько единиц [64,85,171–173]. Для цепочки, изображенной на Рис. 12(г), было показано, что при облучении крайних наночастиц ядро-оболочка фемтосекундным лазером с одной из сторон, симметрия диаграммы направленности разрушается. Это происходит в результате уменьшения диэлектрической проницаемости германиевых (Ge) оболочек, а значит приводит к сдвигу их магнитного резонанса в коротковолновую область. Нарушение симметрии диаграммы направленности выражается в подавлении излучения в сторону немодифицированных наночастиц и усилении в сторону модифицированных наночастиц. Таким образом, данные результаты иллюстрируют потенциал нелинейных гибридных наноантенн типа Уда-Яги для разработки однонаправленных излучателей, способных к сверхбыстрой и обратимой реконфигурации.

Другой подход к управлению оптическими свойствами гибридных наноструктур с помощью высокоинтенсивных световых импульсов связан с необратимой модификацией их формы уже после изготовления [215]. В работе [216] представлен дизайн асимметричных металлодиэлектрических наноструктур и метод их фабрикации с возможностью последующей лазерной обработки, в результате которой металлическая часть наноструктуры меняет свою форму и, как следствие, спектральные характеристики. Такая гибридная наноструктура представляет собой кремниевый конус, на вершине которого находится золотой диск. Стадии литографии обеспечивают специальную форму диэлектрической компоненты для достижения контролируемой модификации металлической компоненты с сохранением магнитного и электрического резонансов Ми, в то время как фемтосекундное лазерное плавление позволяет менять форму золотой наночастицы от диска до чашки или сферы, не влияя на диэлектрическую наночастицу (Рис. 13(а)). Основной механизм, вызывающий изменение формы золотого диска может быть описан через процесс термической деформации [217]. В этом процессе форма поверхности тонкого твердого слоя меняется при нагревании так, чтобы минимизировать поверхностную энергию. Термическая деформация определяется температурой и отношением ширины к толщине нагреваемой наночастицы. В случае асимметричных гибридных наноструктур форма металлической части после фемтосекундной модификации зависит от поглощенной энергии и отношения диаметра к толщине золотого диска.

Поскольку плазмонный резонанс сильно зависит от формы и размеров металлической наночастицы, в которой он возбуждается, контролируемая модификация металлической части гибридной наноантенны (Рис. 13(а)) приводит к его перестройке, в данном случае выражающейся в спектральном сдвиге [216,218,219]. На Рис. 13(б) показано изменение спектра рассеяния и диаграммы направленности в результате фемтосекундного лазерного плавления золотого диска на вершине кремниевого конуса до состояния сферы, сопровождающееся сдвигом резонанса в коротковолновую область спектра и увеличением сечения рассеяния назад на длине волны 600 нм. Также было продемонстрировано, что, используя асимметричные гибридные наноантенны в качестве структурной единицы при создании олигомерных наноструктур, можно модифицировать их коллективный резонансный отклик, в частности, резонанс Фано [85]. На Рис. 13(в) изображены спектры рассеяния гибридных олигомеров, состоящих из перестраиваемых наноантенн, соответствующие различной степени модификации металлических частей наноантенны. В диапазоне длин волн 640-670 нм в спектрах наблюдается резонанс Фано. В работах [220,221] было показано, что этот резонанс имеет магнитную природу, и вызван интерференцией двух мод - спектрально узкой магнитной дипольной моды центрального кремниевого конуса и широкой коллективной магнитной моды кольца из кремниевых конусов. Взаимодействие резонанса Фано диэлектрических компонент гибридного олигомера с плазмонными резонансами металлических компонент выражается в сдвиге и изменении профиля Фано при фемтосекундной лазерной модификации целого олигомера.

## 6. Заключение

Итак, высокие тепловые потери металлических наноструктур ограничивающие их использование во многих современных областях нанофотоники, включая метаповерхности, метаматериалы, нановолноводы, привели к разработке новых устройств, использующих диэлектрические наночастицы с высоким показателем преломления. Такие металлодиэлектрические или гибридные наноструктуры являются в настоящее время авангардом современной науки о свете. Гибридные наноструктуры демонстрируют много интересных с точки зрения практического применения свойств, включая низкие тепловые потери, оптический магнитный резонансный отклик, сильные нелинейно-оптические свойства, обсуждению которых и была посвящена данная обзорная работа. Мы обсудили современное состояние теоретических и экспериментальных исследований гибридных металлодиэлектрических наноантенн и наноструктур на их основе, обладающих способностью избирательно рассеивать световые волны, усиливать и передавать в заданном направлении оптические сигналы, управлять распространением таких сигналов и генерировать оптические гармоники. Кроме того, мы кратко обсудили способы изготовления таких наноструктур на конкретных примерах.

## Благодарности

## Список литературы

integrated cancer imaging and therapy *Nano Lett.* **5** 709–11

[30]  O'Neal D P, Hirsch L R, Halas N J, Payne J D and West J L 2004 Photo-thermal tumor ablation in mice using near infrared-absorbing nanoparticles *Cancer Lett.* **209** 171–6

[31]  Curto A G, Volpe G, Taminiau T H, Kreuzer M P, Quidant R and van Hulst N F 2010 Unidirectional emission of a quantum dot coupled to a nanoantenna. *Science* **329** 930–3

[32]  Esteban R, Teperik T V. and Greffet J J 2010 Optical patch antennas for single photon emission using surface plasmon resonances *Phys. Rev. Lett.* **104** 26802

[33]  Busson M P, Rolly B, Stout B, Bonod N and Bidault S 2012 Accelerated single photon emission from dye molecule-driven nanoantennas assembled on DNA *Nat. Commun.* **3** 962

[34]  Schietinger S, Barth M, Aichele T and Benson O 2009 Plasmon-enhanced single photon emission from a nanoassembled metal - Diamond hybrid structure at room temperature *Nano Lett.* **9** 1694–8

[35]  Lee K, Chen X, Eghlidi H, Renn A, Gotzinger S and Sandoghdar V 2011 A planar dielectric antenna for directional single-photon emission and near-unity collection efficiency *2011 Conf. Lasers Electro-Optics Eur. 12th Eur. Quantum Electron. Conf. CLEO Eur. 2011* **5** 166–9

[36]  Totero Gongora J S, Miroshnichenko A E, Kivshar Y S and Fratalocchi A 2017 Anapole nanolasers for mode-locking and ultrafast pulse generation *Nat. Commun.* **8** 15535

[37]  Sarychev A K and Tartakovsky G 2007 Magnetic plasmonic metamaterials in actively pumped host medium and plasmonic nanolaser *Phys. Rev. B - Condens. Matter Mater. Phys.* **75** 85436

[38]  Zhang C, Lu Y, Ni Y, Li M, Mao L, Liu C, Zhang D, Ming H and Wang P 2015 Plasmonic lasing of nanocavity embedding in metallic nanoantenna array *Nano Lett.* **15** 1382–7

[39]  Suh J Y, Kim C H, Zhou W, Huntington M D, Co D T, Wasielewski M R and Odom T W 2012 Plasmonic bowtie nanolaser arrays *Nano Lett.* **12** 5769–74

photonics *Nat. Photonics* **11** 543–54

[178] Ho J F, Luk'yanchuk B and Zhang J B 2012 Tunable Fano resonances in silver-silica-silver multilayer nanoshells *Appl. Phys. A Mater. Sci. Process.* **107** 133–7

[179] Luk'yanchuk B S, Miroshnichenko a E and Kivshar Y S 2013 Fano resonances and topological optics: an interplay of far- and near-field interference phenomena *J. Opt.* **15** 73001

[180] Wang W, Klots A, Yang Y, Li W, Kravchenko I I, Briggs D P, Bolotin K I and Valentine J 2015 Enhanced absorption in two-dimensional materials via Fano-resonant photonic crystals *Appl. Phys. Lett.* **106** 181104

[181] Yang Y, Wang W, Boulesbaa A, Kravchenko I I, Briggs D P, Puretzky A, Geohegan D and Valentine J 2015 Nonlinear Fano-Resonant Dielectric Metasurfaces *Nano Lett.* **15** 7388–93

[182] Savelev R S, Petrov M I, Sinha R K, Krasnok A E, Belov P A and Kivshar Y S 2015 Fano resonance in chains of dielectric nanoparticles with side-coupled resonator *Proceedings of the International Conference Days on Diffraction 2015, DD 2015* (IEEE) pp 281–4

[183] Shafiei F, Monticone F, Le K Q, Liu X-X, Hartsfield T, Alù A and Li X 2013 A subwavelength plasmonic metamolecule exhibiting magnetic-based optical Fano resonance *Nat. Nanotechnol.* **8** 95–9

[184] Rybin M V., Koshelev K L, Sadrieva Z F, Samusev K B, Bogdanov A A, Limonov M F and Kivshar Y S 2017 High-Q supercavity modes in subwavelength dielectric resonators *ArXiv* 1–5

[185] Metzger B, Hentschel M and Giessen H 2016 Ultrafast Nonlinear Plasmonic Spectroscopy: From Dipole Nanoantennas to Complex Hybrid Plasmonic Structures *ACS Photonics* **3** 1336–50

[186] Li W and Valentine J 2014 Metamaterial perfect absorber based hot electron photodetection *Nano Lett.* **14** 3510–4
51

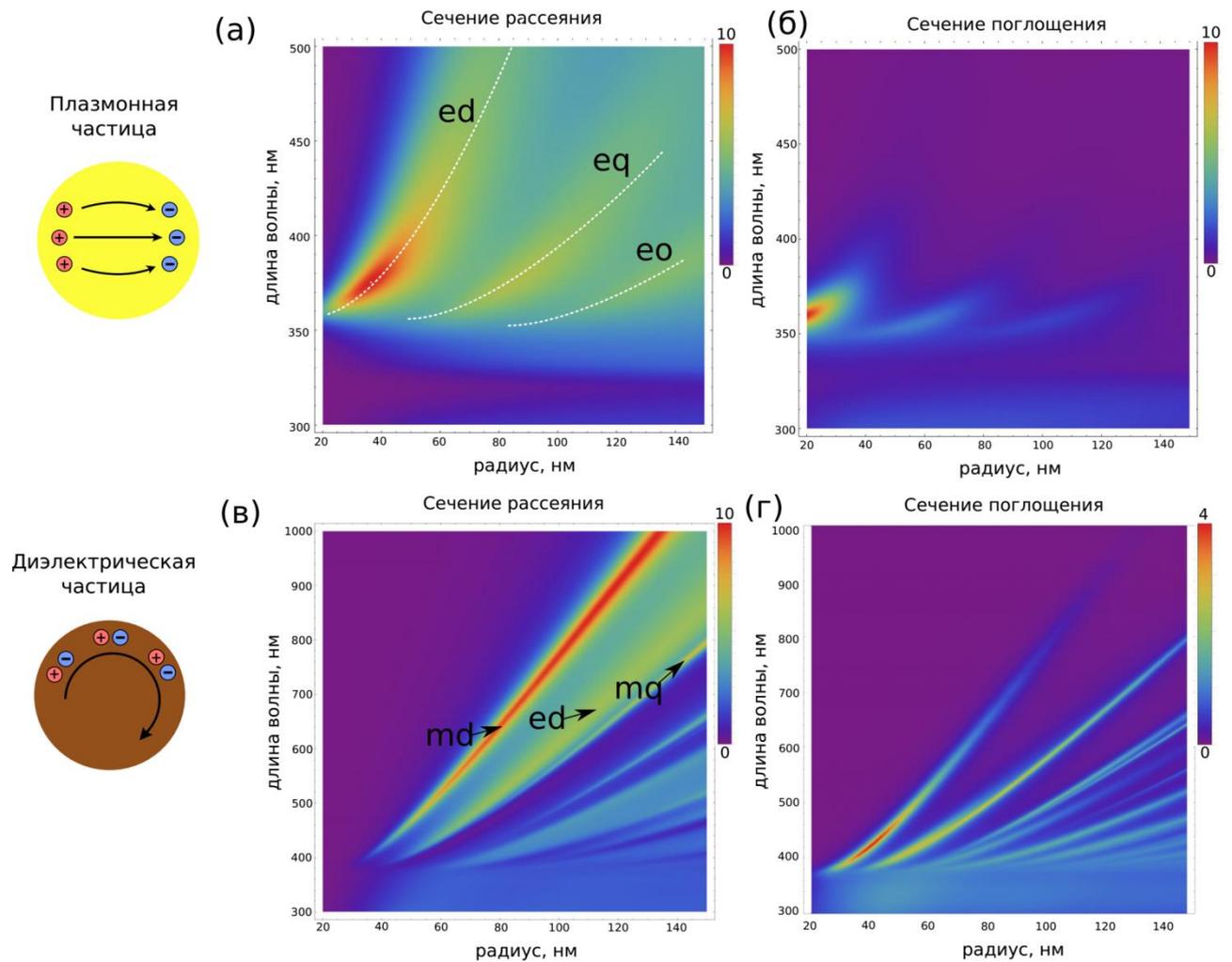

**Рисунок 1.** Оптические свойства плазмонной (Ag) и диэлектрической (Si) наночастиц в свободном пространстве. (а),(в) Нормированное полное сечение рассеяния ($Q_{FS}$) серебряной наночастицы (а) и кремниевой наночастицы (в) в зависимости от их радиуса (R) и длины волны. (б),(г) Нормированное сечение поглощения серебряной наночастицы (б) и кремниевой наночастицы (г) в зависимости от их радиуса и длины волны. Схематические рисунки слева изображают фундаментальные моды плазмонной наночастицы (плазмонный дипольный момент) и диэлектрической наночастицы (магнитный момент Ми).

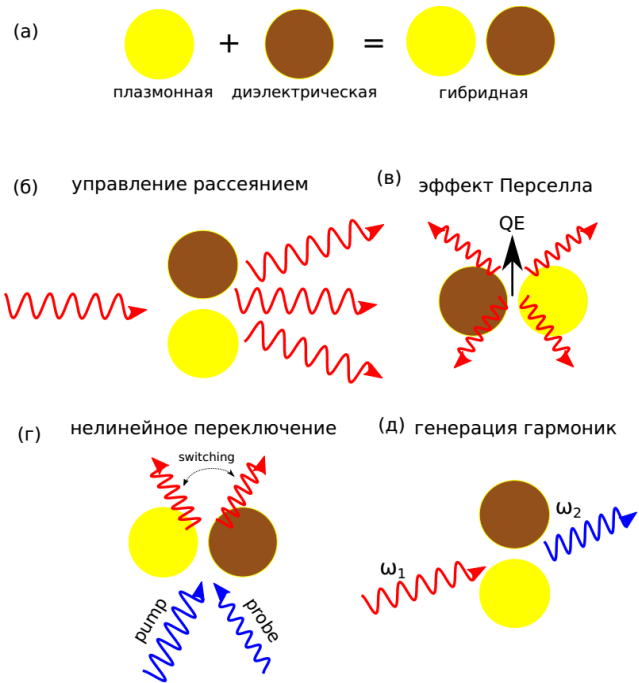

**Рисунок 2.** (а) Суть подхода гибридной нанофотоники: объединение плазмонных и диэлектрических наноструктур в единую систему с целью достижения уникальных оптических свойств. (б)—(д) Различные области практического применения устройств гибридной нанофотоники: (б) линейное управление излучением в дальней зоне, (в) управление скоростью распада возбужденного состояния квантового источника (QE), (г) нелинейное управление рассеяния света при помощи света, (д) генерация новых оптических гармоник.

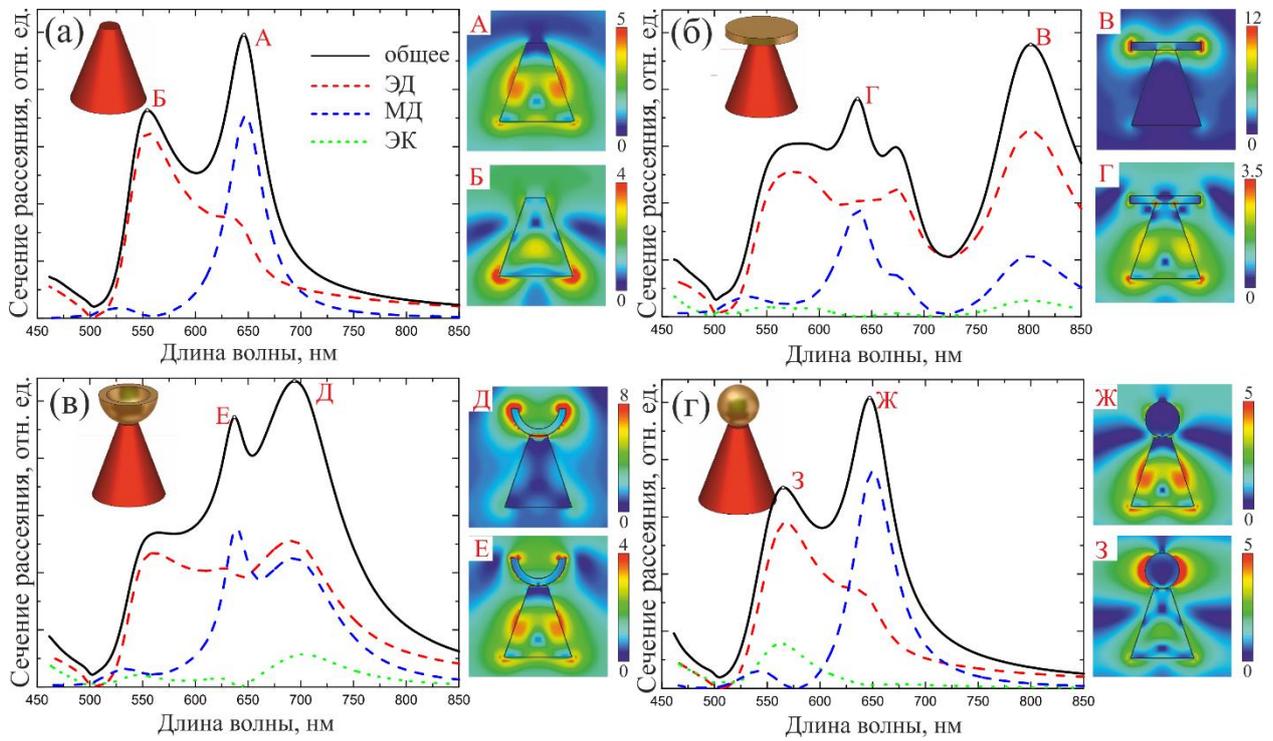

**Рисунок 3.** Зависимости сечения рассеяния диэлектрической (а) и гибридных металл-диэлектрических наноструктур с плазмонными наночастицами в форме диска (б), чашки (в) и сферы (г) в зависимости от длины волны. Пунктирными линиями показаны вклады мультипольных моментов (электрический дипольный – красный, магнитный дипольный – синий, электрический квадрупольный – зеленый). А-З соответствуют распределениям электрического поля на резонансах; значения напряженности электрического поля приведены в В/м. [85]

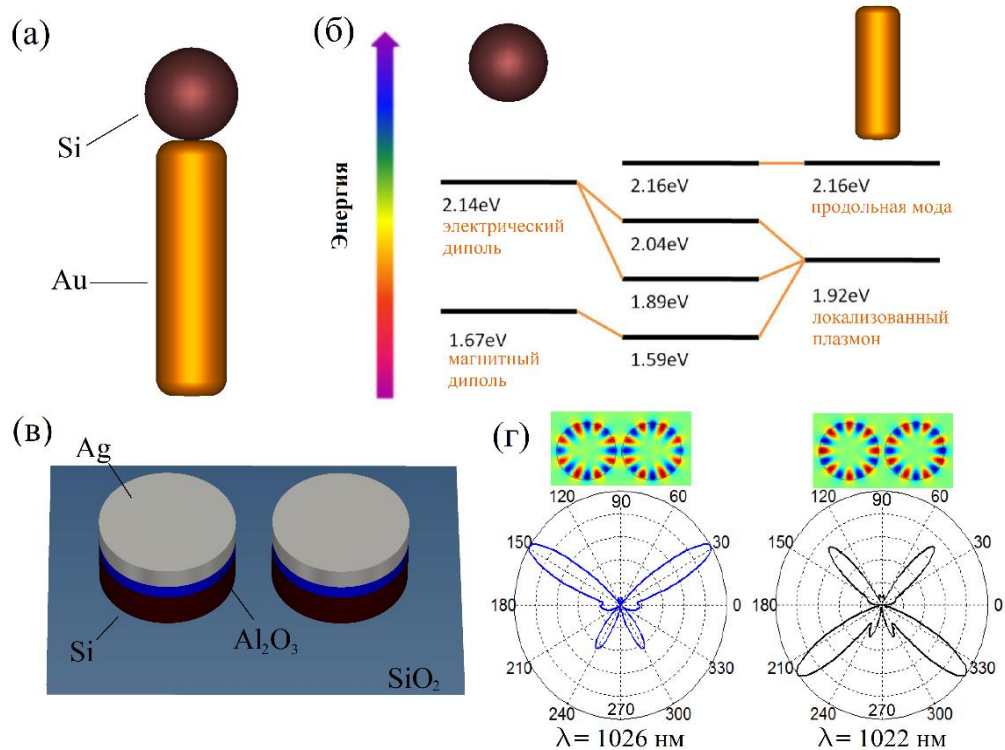

**Рисунок 4.** (а) Схематическая иллюстрация гибридной наноантенны, состоящей из кремниевой (Si) наносферы и золотого (Au) наностержня. (б) Диаграмма каналов рассеяния гибридной наноантенны стержень-сфера. (в) Изображение гибридной димерной наноантенны на стеклянной (SiO2) подложке. (г) Диаграммы направленности димерной наноантенны в E-плоскости, соответствующие симметричной (слева) и антисимметричной (справа) коллективным модам шепчущей галереи. На вставке показаны распределения электрического поля в димерной наноантенне для симметричной и антисимметричной моды шепчущей галереи. [89,96]

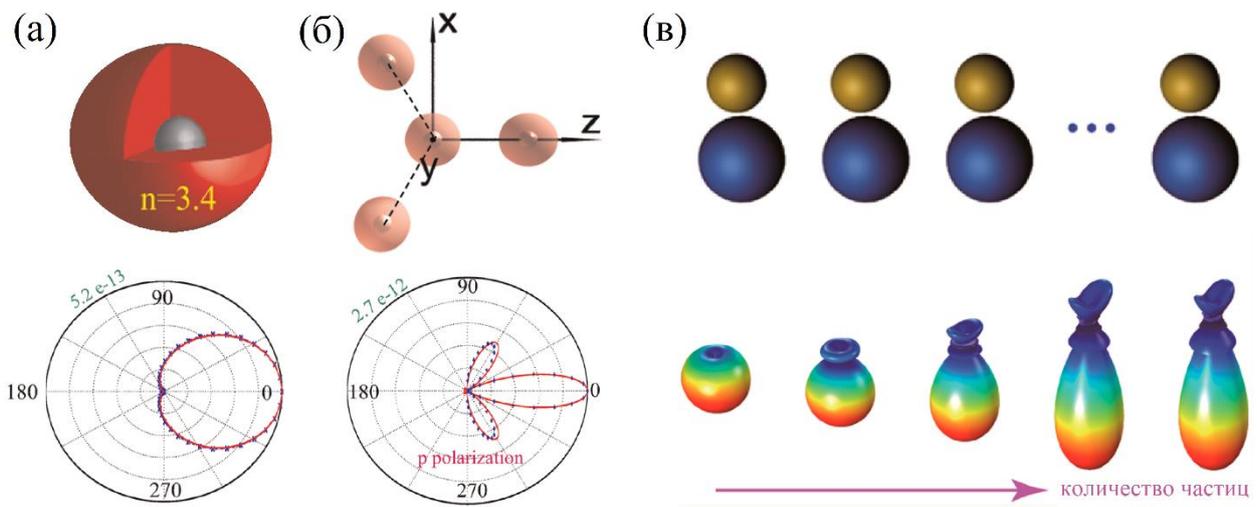

**Рисунок 5.** (а) Схематическая иллюстрация гибридной наноантенны типа ядро-оболочка и (б) составной наноструктуры из таких наноантенн. (в) Изображение цепочки из димерных гибридных сферических наноантенн. Под иллюстрациями приведены диаграммы направленности таких наноантенн. [88,105]

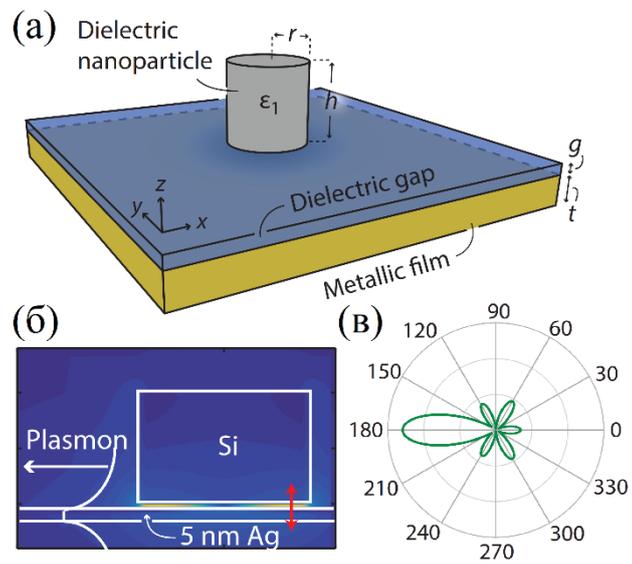

**Рисунок 6.** (а) Изображение и (б) принцип работы гибридной наноантенны, состоящей из диэлектрического диска на серебряной пленке. (в) Диаграмма направленности поверхностного плазмона, возбуждаемого такой наноантенной. [109]

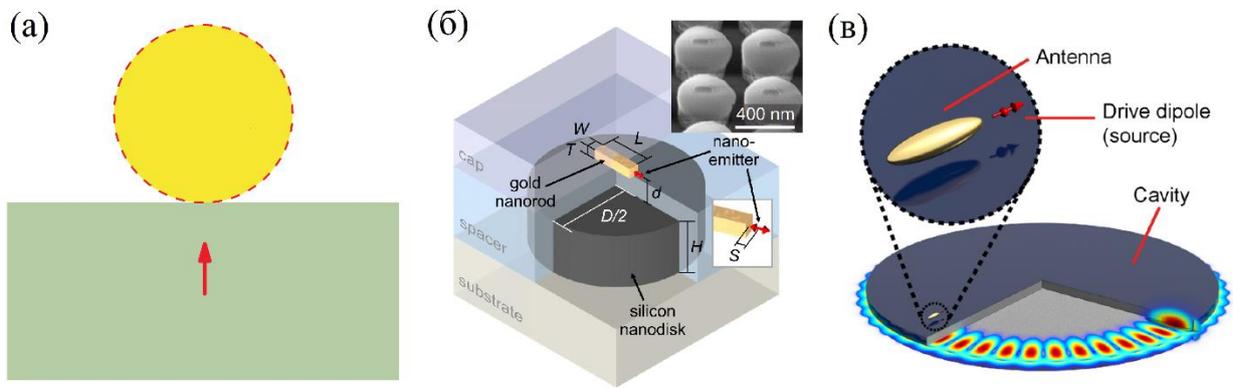

**Рисунок 7.** (а) Схематическое изображение гибридной наноантенны, включающей золотую наночастицу и планарную высокоиндексную подложку. (б) Иллюстрация гибридной наноантенны типа нанодиск-наностержень. Красными стрелками показаны квантовые источники света. (в) Гибридная система на основе дискового микрорезонатора и металлической наноантенны. [87,111,112]

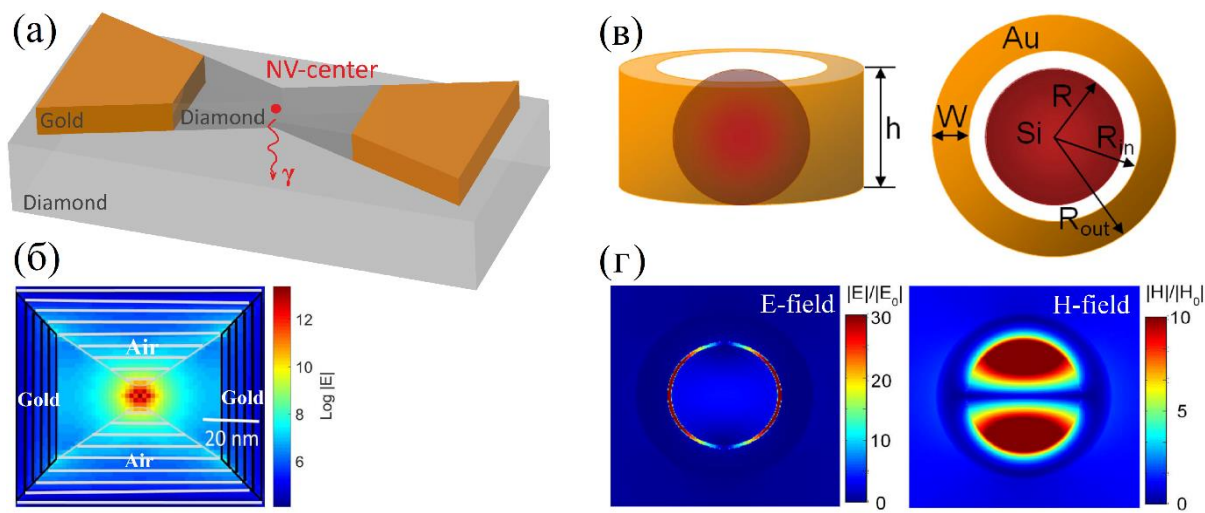

**Рисунок 8.** (а) Изображение гибридной наноантенны типа галстук-бабочка, состоящая из золота и алмаза с внедренными NV-центрами. (б) Профиль в горизонтальном сечении электрической дипольной моды, локализованной в гибридной наноантенне (белыми штрихами показана область, занимаемая воздухом, черными штрихами – золотыми трапециями). (в) Схематическая иллюстрация металлического кольца и диэлектрической сферы, объединенных в гибридную наноантенну. (г) Распределения электрического и магнитного поля в такой наноантенне на длине волны резонанса Фано. [113,167]

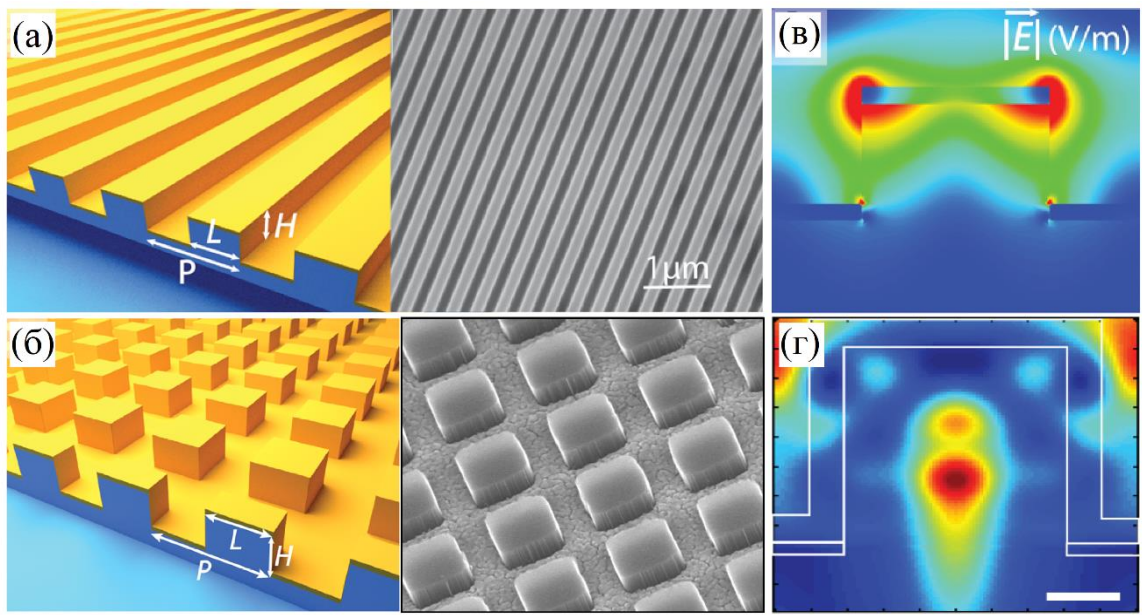

**Рисунок 9.** Схематическое и SEM изображения (а) одномерной решетки и (б) метаповерхности из гибридных наноантенн. металлического кольца и диэлектрической сферы, объединенных в гибридную наноантенну. Распределения электрических полей в (в) структруной ячейке одномерной решетки и (г) гибридной наноантенне. [186,187]

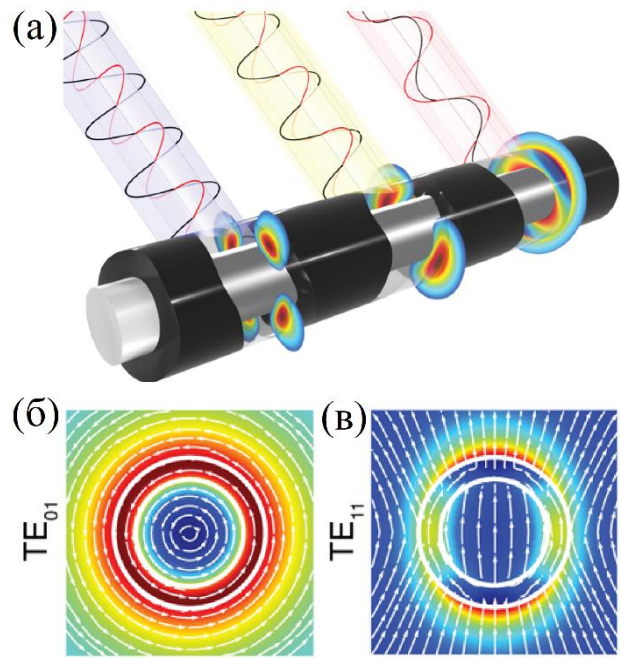

**Рисунок 10.** (а) Иллюстрация гибридной наноструктуры типа ядро-оболочка, состоящей из серебряного стержня, покрытого полупроводниковой поглощающей оболочкой. Распределения электрических полей на длинах волн (б) $TE_{01}$ и (в) $TE_{11}$ резонансов гибридной наноструктуры. [188]

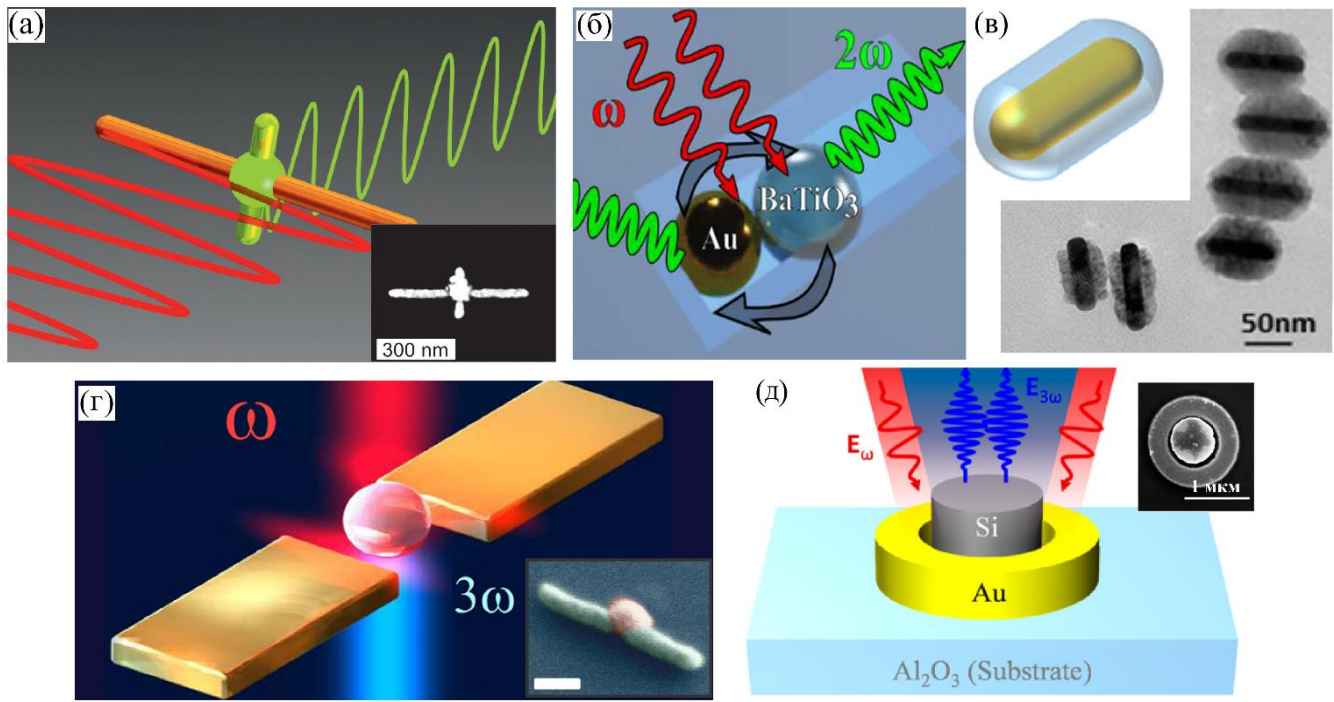

**Рисунок 11.** (а) Иллюстрация гибридной наноантенны для эффективной генерации второй гармоники и принципов ее работы. (б) Гибридный нанодимер, состоящий из Au и $BaTiO_3$ сферических наночастиц. (в) Изображения гибридной Au-$Cu_2O$ наноантенны типа ядро-оболочка. (г) Иллюстрация гибридной дипольной наноантенны с активной ITO наночастицей. (д) Схема работы гибридной наноструктуры кольцо-диск на сапфировой ($Al_2O_3$) подложке; на вставке показано изображение такой наноструктуры, полученной сканирующей электронной микроскопией. [193,195–197]

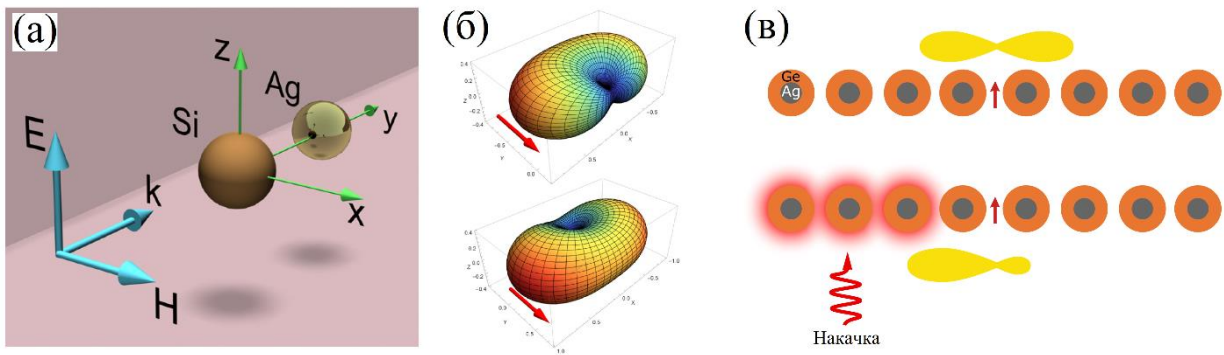

**Рисунок 12**. (а) Изображение нелинейной гибридной димерной Si-Ag наноантенны. (б) Диаграммы направленности такой наноантенны при облучении световыми волнами низкой интенсивности (вверху) и высокой интенсивности (внизу); красными стрелками показано направление волнового вектора световой волны. (с) Схема работы нелинейной цепочки гибридных наночастиц типа ядро-оболочка; желтым показаны диаграммы направленности квантового источника света, помещенного в центр цепочки. [6,210]

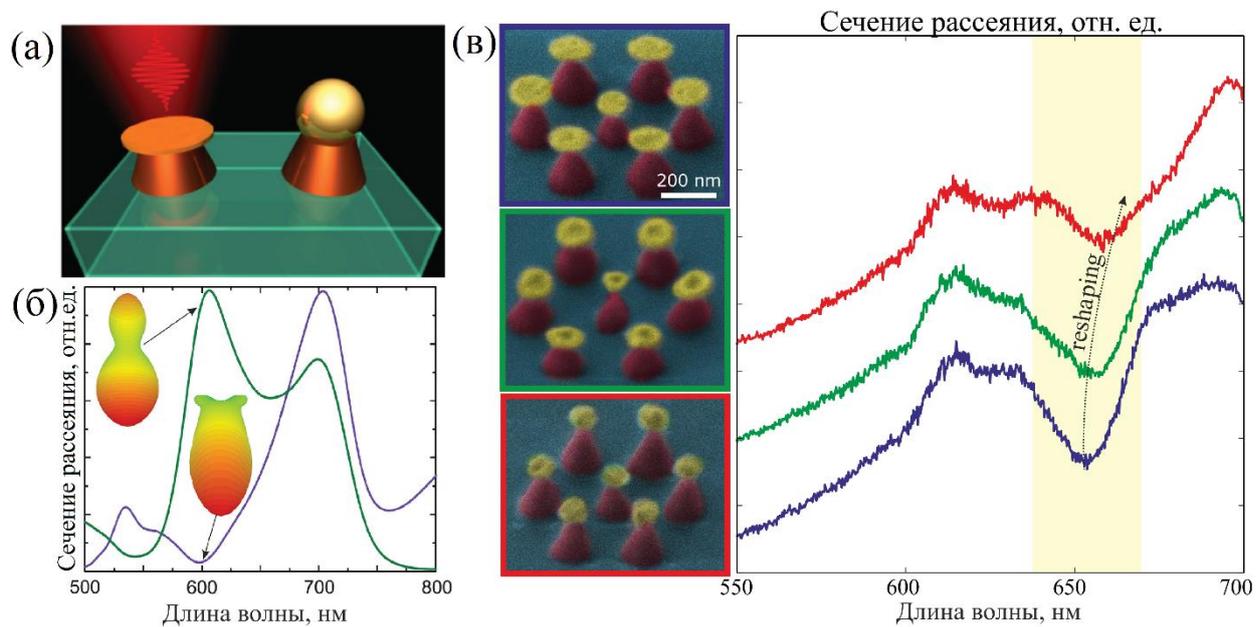

**Рисунок 13**. (а) Иллюстрация концепции перестраиваемых асимметричных гибридных наноантенн. (б) Спектры рассеяния гибридных наноантенн с металлической наночастицей в форме диска (синяя кривая) и сферы (зеленая кривая). (в) СЭМ-изображения гибридных олигомеров с различной степенью модификации металлических наночастиц и их спетры рассеяния. [85,216]